\documentclass[%
 reprint,
 amsmath,amssymb,
 aps,
]{revtex4-2}

\usepackage{graphicx}
\usepackage{dcolumn}
\usepackage{bm}
\usepackage{hyperref}

\begin{document}

        
\title{Electron Energy Spectra and \mbox{\textit{e-e}}~Bremsstrahlung from Anisotropic Electron Distributions in Extreme Solar Flares}

\author{\firstname{I.~D.}~\surname{Oparin}}
\email{ivandmit@gmail.com}
\affiliation{%
	Ioffe Institute, Russian Academy of Sciences, St.~Petersburg, 194021 Russia
}%
\author{\firstname{Yu.~E.}~\surname{Charikov}}
\affiliation{%
	Ioffe Institute, Russian Academy of Sciences, St.~Petersburg, 194021 Russia
}%

\author{\firstname{E.~P.}~\surname{Ovchinnikova}}
\affiliation{%
	Ioffe Institute, Russian Academy of Sciences, St.~Petersburg, 194021 Russia
}%

\author{\firstname{A.~N.}~\surname{Shabalin}}
\affiliation{%
	Ioffe Institute, Russian Academy of Sciences, St.~Petersburg, 194021 Russia
}%

\date{\today}

\begin{abstract}
	We study the electron energy spectra of two powerful solar flares SOL2003-10-28T 11:06:16 (\textit{GOES} class~X17.2) and SOL2002-07-23T00:18:16 (X4.8) on the basis of HXR and gamma-ray spectra obtained from \textit{RHESSI} data. Electron-electron (\mbox{\textit{e-e}}) bremsstrahlung makes a significant contribution to the X-ray flux at energies above 500~keV. In X-ray flux calculations different electron pitch-angle distributions were considered: quasi-transverse and quasi-longitudinal. It was shown that breaks in hard X-ray spectra occur on energies $\epsilon\approx100$~keV with photon spectral-index difference $\Delta\gamma\ge 1$  in the case of quasi-longitudinal anisotropy and hard power-law electron energy spectra. We deduce the parameters of non-thermal electrons energy spectra from \textit{RHESSI} data for the extreme solar flares of 2003 October 28 and 2002 July 23. After including e-e bremsstrahlung, the non-thermal electrons energy spectra obtained in the thick-target model can be fitted using power-law with spectral index $\delta\approx 5$. For the 2002 July 23 flare, there is a pronounced asymmetry towards the southern footpoint, starting  from energies above 100 keV. The separation of HXR sources at energies above 100~keV may be related to a feature of relativistic electron transport in flaring loops. 
\end{abstract}

\maketitle


\section{INTRODUCTION}

Hard X-ray (HXR) emission in solar flares is produced by bremsstrahlung from accelerated electrons with energies 
$E>20$~keV located along the flaring loop (e.g.~\cite{Asch}). For the majority of events, the non-thermal radiation with power-law spectrum is observed in the energy range from 20~keV to 150~keV. However, for the extreme powerful flare events (\textit{GOES} class X) bremsstrahlung continuum in gamma rays can be detected up to the energies about tens of MeV, which indicates the acceleration of electrons to relativistic energies.

HXR flux in the energy range between 20 and 150~keV is produced mostly through dipole electron-ion (\textit{e-i}) bremsstrahlung. In case of electron-electron scattering, the bremsstrahlung is quadrupole, because for this system the second derivative of the dipole moment is zero. The total emitted energy turns out $(v/c)^2$ times less comparing to scattering on an ion or nucleus at rest \cite{Zhel}. Electron-electron (\mbox{\textit{e-e}}) bremsstrahlung becomes efficient for the relativistic electrons with energies $E>511$~keV and makes a comparable contribution to HXR flux \cite{Hg75}. It has been shown by Kontar et al. \cite{Kontar} that \mbox{\textit{e-e}} bremsstrahlung should be taken into account while recovering electron flux spectra on energies above 300~keV in assumption that radiating electrons are isotropic.

The observations of coronal HXR and microwave sources and kinetic simulations of electron transport in flaring loops point out the existence of anisotropic electron pitch-angle distributions \cite{Rezn,Char}. To obtain the HXR parameters such as flux, directivity, and degree of linear polarization these pitch-angle distributions have to be convoluted with anisotropic relativistic \mbox{\textit{e-e}} and \mbox{\textit{e-i}} bremsstrahlung cross-sections. Also, it is noteworthy that the higher the electron energy the greater the longitudinal anisotropy of the bremsstrahlung cross-sections. Thus, the degree of anisotropy of the electron distribution function can be estimated from the measurements of the HXR flux in a broad energy range. The estimation of \mbox{\textit{e-e}} bremsstrahlung contribution from anisotropic electron distributions without considering the dependence on the viewing angle was made previously by \cite{Hg75}. In the present paper, we consider the marginal cases of pitch-angle distributions to study the influence on HXR flux. It should be noted that direct anisotropy diagnostics techniques such as polarization or directivity measurements are not applicable in the high-energy range ($>100$~keV), due to the falling of linear polarization degree with the growth of photon energy \cite{Zharkova}, which extends the exposure time up to dozens of minutes so polarization can change appreciably.

The data from the \textit{RHESSI} spectrometer \cite{Lin02} with uniquely high energy (about 3~keV above 100~keV) and temporal resolution (4~seconds) allows to obtain electron energy spectra from observed HXR flux through solving an inverse problem. It is also possible to reconstruct the images of HXR and gamma-ray sources at energies above 100~keV in powerful flares with spatial resolution ~ 35$''$\cite{Hur03}. HXR and gamma-ray imaging in the energy range corresponding to the bremsstrahlung continuum reveals the localization of high-energy electrons.

HXR imaging spectroscopy eventually yield the spatial distribution of HXR and gamma ray brightness, the parameters of accelerated electrons beam and plasma in the HXR and gamma ray emission sites, and therefore, could place constraints on the model electron distribution functions and acceleration mechanism.

In this article we also examine the structure of HXR and gamma-ray sources in extreme solar flares SOL2003-10-28~(X17.2) and SOL2002-07-23~(X4.8) and special attention will be paid to focus on the influence of \mbox{\textit{e-e}} bremsstrahlung in the reconstruction of energy spectra of relativistic electrons taking into account the pitch-angle anisotropy.

\section{ELECTRON-ELECTRON BREMSSTRAHLUNG}

Bremsstrahlung cross-section for \mbox{\textit{e-i}} scattering is obtained in assumption that scattering center is motionless, however, in the \mbox{\textit{e-e}} scattering the recoil of target electron can not be omitted. In view of that, the energy of the radiating electron is a function of the angle between electron and photon momentum vectors and also photon energy. Furthermore, on account of electrons are identical particles the cross-section of \mbox{\textit{e-e}} bremsstrahlung allows for the exchange interaction between electrons. The covariant expression for \mbox{\textit{e-e}} bremsstrahlung cross-section was derived by E.~Haug \cite{Hg75a}:

\begin{equation*}
	\frac{d^2\sigma_{ee}}{d\epsilon d\Omega_k}=\frac{\alpha_{FS} r_0^2}{\pi}\frac{k}{w\rho}\sqrt{\frac{\rho^2 - 4}{w^2 -4}}\frac{1}{\pi}\int A d\Omega_{p_1'}
\end{equation*}

where $w^2 = (\underline{p_1}+\underline{p_2})^2$, $\rho^2 = (\underline{p_1}'+\underline{p_2}')^2$ are kinematic invariants which depend on electrons 4-momentum before and after scattering, $A$ is scattering amplitude, which is integrated on the electron scattering solid angle $d\Omega_{p_1'}$, $\alpha_{FS}$ is the fine structure constant, $r_0$ is the classical electron radius and $k=h\nu/m_ec^2$.

It is convenient to study the scattering of relativistic electrons beam on the thermal plasma electrons in the laboratory reference frame, assuming plasma electrons motionless: $p_2 = 0$.

Quantitative comparison of angle-integrated \mbox{\textit{e-e}} and \mbox{\textit{e-i}} bremsstrahlung cross sections reveals that for electron energies $E>1$~MeV \mbox{\textit{e-e}} bremsstrahlung cross-section exceeds \mbox{\textit{e-i}} cross-section, hence \mbox{\textit{e-e}} bremsstrahlung contribution to the total HXR flux is determined by the number of high-energy electrons, i.e. by electron energy spectra hardness. Thus \mbox{\textit{e-e}} bremsstrahlung becomes principal HXR radiation mechanism for mildly relativistic electrons.

\begin{figure*}[htbp]
	\input{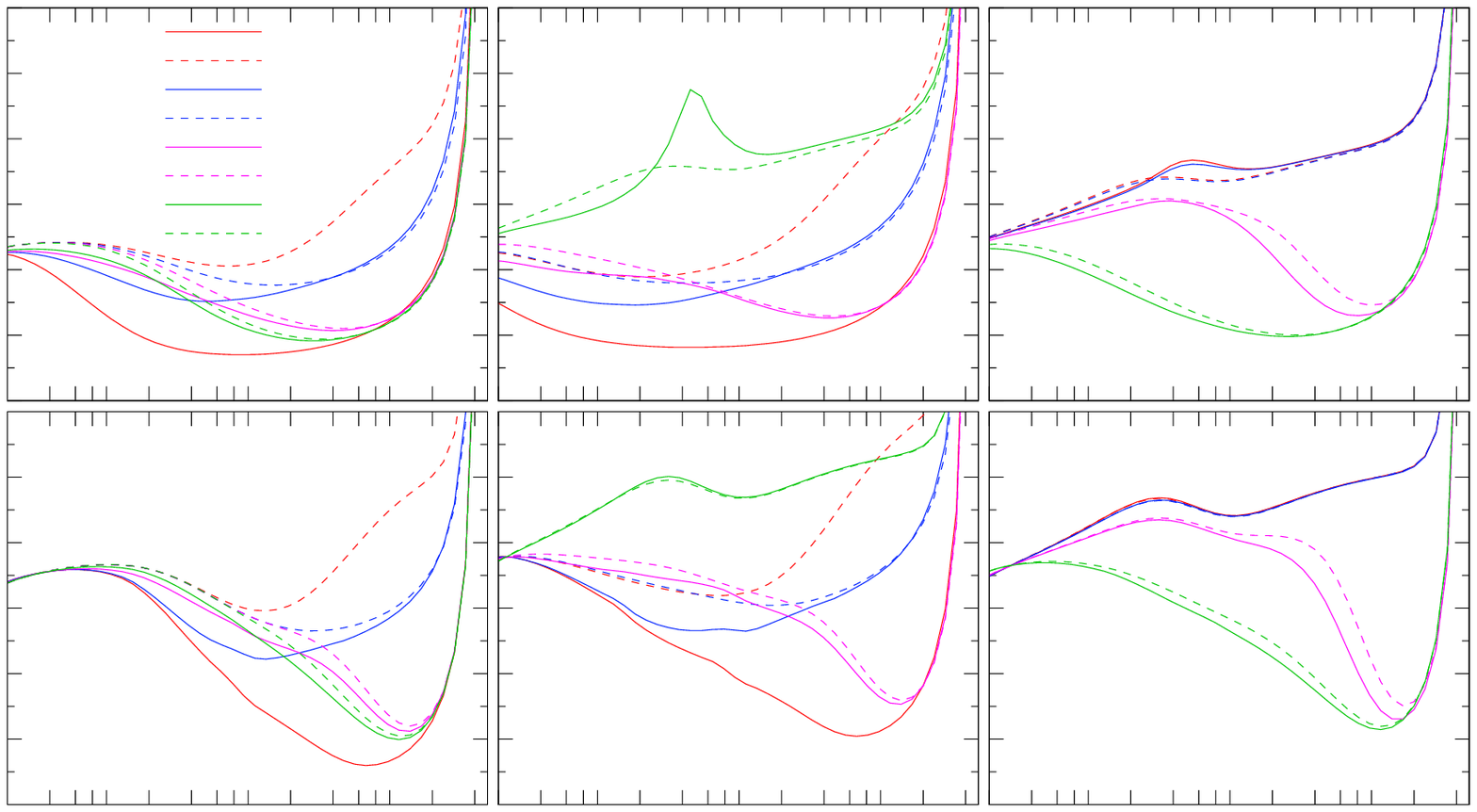}
	\caption{\label{fig:wide}HXR energy spectra for a power-law electron energy spectra characterized by spectral indices $\delta = 2.5$ and $\delta = 5$ for different viewing angles $\alpha = 0,~10,~30$, and 90 degrees. For each viewing angle, the lower curve corresponds to the flux on the e-i bremsstrahlung, and the upper curve corresponds to the total flux (\mbox{\textit{e-i}} + \mbox{\textit{e-e}}). The choice of maximal energy in the electron spectrum $E_{max}=50$~MeV leads to a cut-off in the HXR spectrum on $E_{max} =\epsilon_{max}$.}
\end{figure*}

\section{COMPUTATION OF HARD X-RAY FLUX}

The diagnostics of high energy electrons anisotropy from solar HXR flux measurements is a nontrivial problem (see \cite{Massone}), particularly for relativistic electron energies. It is necessary to calculate HXR and gamma-ray flux in the energy range above 100 keV using electron distribution functions and taking into consideration \mbox{\textit{e-i}} and \mbox{\textit{e-e}} bremsstrahlung. Accelerated electrons flux is expressed through distribution function $f({v},{r},t)$ as $F(E,\Omega,r,t)dE = {v}f({v},{r},t)$.  We also assume that the distribution function can be factorized into an energetic and angular part which is azimuthally symmetric, i.e. we represent $F(E,r=r_0,\Omega,t=t_0)\equiv F(E,\Omega)=N(E)g(\theta,\phi),=AE^{-\delta}g(\theta)$, where $\phi,\theta$ are the azimuthal and pitch angles of relativistic electrons. 

HXR and gamma-ray flux produced by \mbox{\textit{e-i}} and \mbox{\textit{e-e}} bremsstrahlung (first and second term respectively) is defined as follows:

\begin{multline}
	I_{ep+ee}(\epsilon,\alpha)=\\
	\frac{\overline{n}V}{4\pi R^2}\int_{\epsilon}^{\infty}dE\int_{-1}^{+1}d\cos\theta \int_{0}^{2\pi}d\phi N(E)\times \\g(\theta) v(E) \frac{d^2\sigma_{ep}}{d\epsilon d\Omega_k}(E,\epsilon,\psi)+\\\frac{\overline{n}V}{4\pi R^2}\int_{-1}^{+1}d\cos\theta\int_{0}^{2\pi}d\phi\int_{E_0(\psi(\alpha),\epsilon)}^{\infty}dE N(E)\times\\g(\theta)v(E)\frac{d^2\sigma_{ee}}{d\epsilon d\Omega_k}(E,\epsilon,\psi)
\end{multline}

Here $\alpha$ is the viewing angle (angle between wave vector $\mathbf{k}$ and magnetic field $\mathbf{B}$ at the moment of HXR photon emission, measured from solar center to limb), $\epsilon$ is the photon energy, $V$ is the HXR emitting volume, $n_p, n_e$ are the plasma densities of thermal ions and electrons in volume $V$,  $N(E)$ is the electron energy spectra, taken in the form of the power-law: $N(E)=AE^{-\delta}$, $v(E)$  is the relativistic electron velocity, $\psi$ is the angle between initial momentum vector of electron and photon wave vector (scattering angle), $R$ is 1 astronomical unit. Relativistic \mbox{\textit{e-i}} bremsstrahlung cross-section $\frac{d^2\sigma_{ee}}{d\epsilon d\Omega_k}(E,\epsilon,\psi)$, is given by \cite{GHull}.

Because of target electron recoil the lower integration limit (minimal energy of radiating electron) is related to the scattering angle and photon energy \cite{Hg75}:
$$ E_0(\psi,\epsilon)=\epsilon\frac{2(1+k)+\cos\psi(k\cos\psi-\sqrt{k^2\cos^2\psi+4k})}{1-k^2+k\cos\psi\sqrt{k^2\cos^2\psi+4k}}
$$  
The angles $\psi,\alpha,\theta,\phi$, are interconnected: $\cos\psi={(\mathbf{k}\mathbf{p_1})}/{kp_1}=\sin\alpha\sin\theta\cos\phi+\cos\alpha\cos\theta$. We also introduce the local spectral slope $\gamma$, approximating the HXR spectrum in the local energy interval by a power law:

\begin{equation*}
	\gamma(\epsilon)=-\frac{\epsilon}{I(\epsilon)}\frac{dI}{d\epsilon}=-\frac{d\log I}{d \log \epsilon}
\end{equation*}

We examine the HXR flux parameters for marginal cases of pitch-angle distribution: isotropic $g(\theta)=1$, quasi-longitudinal $g(\theta)=\cos^8\theta$  and quasi-transversal $g(\theta)=\sin^8\theta$. The quasi-longitudinal distribution can occur if electrons  precipitate to the chromosphere through the loss-cone, on the contrary, a quasi-transverse electron distribution can occur in the looptops, provided that electrons with a low pitch angle fall into the loss-cone, and electrons with large pitch angles are mirroring in a converging magnetic field.

For isotropic electron distribution $g(\theta)=1$ HXR flux and the local photon spectral index depend on the viewing angle $\alpha$ (Fig.~1, left). As can be seen in figure 1, for the hard electron energy spectra with spectral index $\delta = 2.5$ and isotropic distribution, the local photon spectral index varies from 3.5 to 3.0 in the energy range below 100~keV and viewing angles above $10^\circ$. We emphasize that in this case the spectral index of total HXR flux (\mbox{\textit{e-i}} and \mbox{\textit{e-e}} bremsstrahlung) is virtually unchanged from spectral index considering \mbox{\textit{e-i}} bremsstrahlung only, which is true for the non-relativistic energy range. Especially note the case for small viewing angles ($\alpha<10^\circ$, events in the disc center), when the difference in spectral indices for \mbox{\textit{e-i}} bremsstrahlung and total bremsstrahlung flux $\Delta\gamma = |\gamma_{e-i}-\gamma_{e-e+e-i}|\le~1$ throughout the entire energy range. Conversely for the large viewing angles ($\alpha\sim90^\circ$, limb events) the index difference $\Delta\gamma=0$ over the entire energy range, that is, the contribution of \mbox{\textit{e-e}} bremsstrahlung is negligible. When electron energy spectra soften to $\delta = 5$ the local HXR spectral index $\gamma$ varies within 0.5 in the broad energy range. A notable distinction in $\gamma$ for \mbox{\textit{e-i}} and \textit{e-i+e-e} can be also pointed out for viewing angles $\alpha<10^\circ$ and photon energies 100~keV. 

\begin{figure*}
	\includegraphics[scale=0.45]{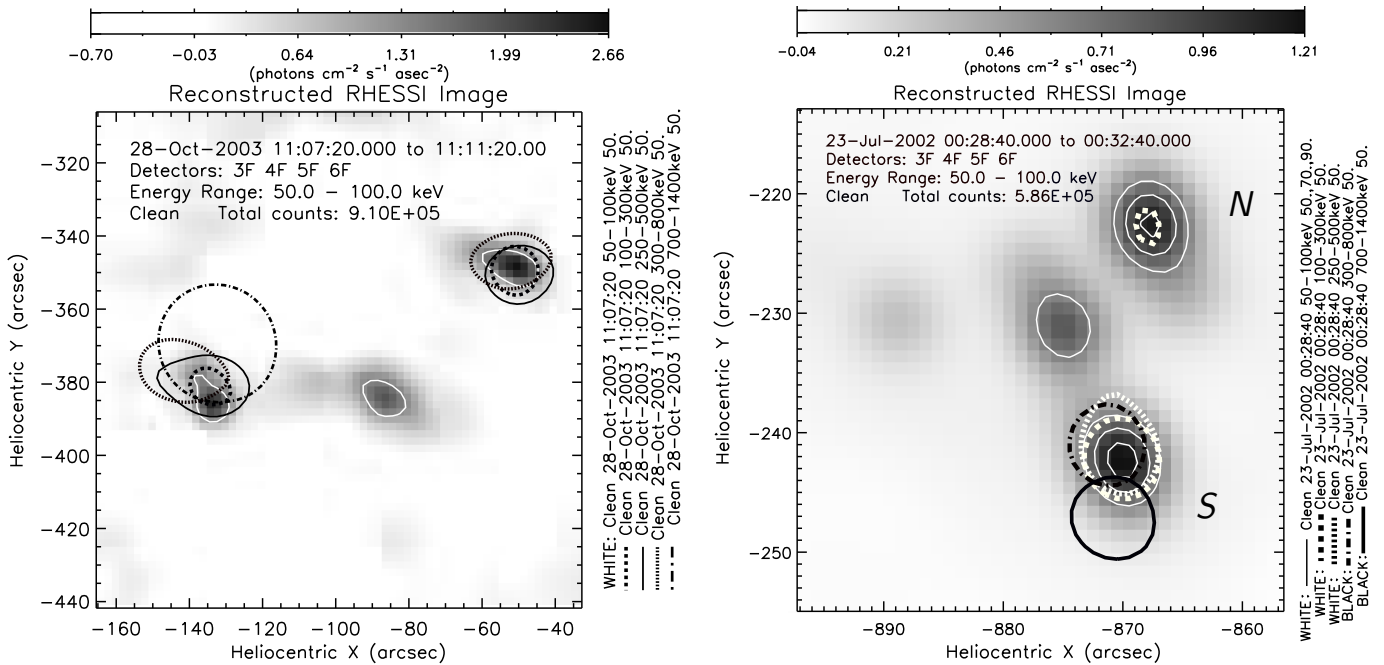}
	\caption{\label{fig:2wide} X-ray and gamma-ray images of 28.10.2003 (left) and 23.07.2002 (right) flares, reconstructed from \textit{RHESSI} data using \textit{CLEAN} algorithm in the energy range from 50 to 100~keV. The
		image superimposed contours in energy bands 50-100~keV, 100-300~keV, 250-500~keV, 300-800~keV, and 700-1400~keV.}
\end{figure*}

If high energy  electrons are longitudinally distributed in HXR source $g(\theta)=\cos^8\theta$  (Fig. 1, center) with hard energy spectra $\delta = 2.5$, the pitch angle anisotropy causes differences in \mbox{\textit{e-i}} and \textit{e-i+e-e} photon spectral indices $\Delta\gamma>1$ for photon energies 100~keV and viewing angles $\alpha$ from $0^\circ$ to $10^\circ$ which correspond to the center of the solar disc. Hence, \mbox{\textit{e-e}} bremsstrahlung from relativistic electrons with hard energy spectra $\delta=2.5$ and observation angles $\alpha<10^\circ$ determines the directivity of HXR and gamma radiation. As the viewing angle increases, the distinction in local spectral indices decreases. In this case, we also remark a peak on photon energy $\epsilon$ about 500~keV and $\alpha<90^\circ$. This peak is caused by the presence of high-energy limit in photon energy in \mbox{\textit{e-e}} bremsstrahlung, which depends on the electron energy and scattering angle \cite{Hg75}. That is, that photon energy has maximum for the specified parameters, and the generation of HXR by \mbox{\textit{e-e}} bremsstrahlung above 500~keV has ceased. For the soft electron energy spectra with spectral index $\delta=5$ the influence of \mbox{\textit{e-e}} bremsstrahlung is marginal due to the low number density of high energetic electrons. Also, the dependence of HXR spectra on viewing angle is reducing and difference in photon spectral indices exceeds 1 only for photon energy range above 1~MeV (Fig. 1, right), where the electrons number density is small for spectral index $\delta=5$.

The examination of local HXR spectra allows us to conclude that break in the photon energy spectra below 100~keV, caused by the growth of \mbox{\textit{e-e}} bremsstrahlung flux, takes place for quasi-longitudinal pitch angle distribution of radiating electrons and electron spectral index $\delta = 2.5$ (Fig. 1, center).

There are virtually no differences in photon spectral indices with and without including \mbox{\textit{e-e}} bremsstrahlung for quasi-transversal pitch angle distribution $g(\theta)=\sin^8\theta$ (Fig. 2, right). The existing dependence on the observation angle is conditioned by the transversal component of \mbox{\textit{e-e}} bremsstrahlung cross-section.

\section{HARD X-RAY IMAGING OF 28.10.2003 AND 23.07.2002 FLARES}

In the paper we analyze HXR data from two extremely powerful solar flares SOL2003-10-28T~11:06:16 (\textit{GOES} class X17.2) and SOL2002-07-23T~00:18:16 (\textit{GOES}~X4.8). Figure~2 shows reconstructed X-ray images of these flares in various energy bands: 50-100~keV, 100-300~keV, 250-500~keV, 300-800~keV, 700-1400~keV according to \textit{RHESSI} observations \cite{Hur02}. The contours in energy bands up to 1.4~MeV are superimposed on the image in 50-100~keV energy range (white solid). For each flare it is possible to pick out three local sources in the 50-100~keV energy band, the local source in the middle coinciding with the thermal source in the  12~--~25~keV energy band and can be identified with the looptop of the magnetic structure, and two other sources located in the footpoints of magnetic loops. The space location of non-thermal X-ray sources in 28 October 2003 flare is symmetrical (Fig. 2, left) for all energies, while for the 23 July 2002 flare (Fig. 2, right), there is a pronounced asymmetry towards the southern footpoint, starting from energies above 100~keV. We also point out the lack of gamma radiation in 0.7~--~1.4~MeV energy band in the northern footpoint, whilst there is such source in a southern footpoint. The difference in HXR brightness at energies above 100~keV and absence of northern gamma-ray source may be related to a feature of relativistic electron transport in flaring loops caused by the coupling of the angular and energy parts of the electron distribution function. Further research requires kinetic modeling of electron transport with a specific distribution function.

Gamma radiation sources, which are localized in footpoints with high plasma density, usually assume bremsstrahlung in a thick-target model. We also mention that Coulomb collisions in a dense plasma can efficiently isotropize pitch angle distribution of accelerated electrons. Thus, pitch angle distribution of accelerated electrons in gamma ray sources could be considered as isotropic.

\section{RECOVERING ELECTRON ENERGY SPECTRA}

To reconstruct the accelerated electrons flux from the observations of HXR and gamma-rays in flares on October 28, 2003 and July 23, 2002 we applied a direct approximation method (forward-fitting) consisting in fitting the parameters of the electron energy spectrum to the observed HXR spectrum using the chi-square test. In the thick-target model \cite{Hol11}, the formula for HXR flux by \mbox{\textit{e-i}} and \mbox{\textit{e-e}} mechanisms for isotropic electron distribution has the form:

\begin{multline}
	I(\epsilon) = \\
	\frac{n_p N_b A}{4\pi R^2}\int_{\epsilon}^{E_{max}}F(E)v\left[\int_{\epsilon}^{E}\frac{d\sigma_{ep}}{dk}(\mathcal{E},\varepsilon)\frac{1}{d\mathcal{E}/dt}d\mathcal{E}\right]dE \\
	+ \frac{n_e N_b A}{4\pi R^2}\int_{E_0(\epsilon)}^{E_{max}}F(E)v\left[\int_{E_0(\epsilon)}^{E}\frac{d\sigma_{ee}}{dk}(\mathcal{E},\epsilon)\frac{1}{d\mathcal{E}/dt}\right]dE		
\end{multline}

Here $F(E)=KE^{-\delta}$ is the electron flux density [electron s$^{-1}$ keV$^{-1}$], $A$ is the area of the X-ray source, $N_b$ is the accelerated electrons number density, $n_p=n_e$ is the thermal plasma number density. The relativistic formulae for the \mbox{\textit{e-i}} and \mbox{\textit{e-e}} bremsstrahlung cross-sections are taken from \cite{Hg97,Hg98}, respectively. Relation for minimal electron energy \cite{Hg75}: $$E_0(\epsilon)=\epsilon\frac{2+3k-\sqrt{k^2+4k}}{1-k^2+k\sqrt{k^2+4k}},$$ energy loss in Coulomb collisions: $dE/dt=4\pi r_0^2(mc^2)nc\ln\Lambda/\beta$ \cite{LP}. We implemented the algorithm for approximating the HXR spectra using the formula (2) based on the thick-target model in the \textit{OSPEX} spectrometric package \cite{Hol11}.

Fig. 3 shows the HXR spectra calculated in the thick target model with (EPB+EEB) and without (EPB) \mbox{\textit{e-e}} bremsstrahlung for power-law electron energy spectra characterized by $\delta = 2,~3,~5$. From the analysis of the spectra (Fig. 3), it follows that in the HXR spectrum calculated using (2) for power law electron spectra with the indices $\delta = 2,~3,~5$, when taking into account the \mbox{\textit{e-e}} bremsstrahlung, a pronounced spectral break occurs at energy ~ 400 keV for the soft electron spectrum with the index $\delta = 5$ and an isotropic pitch angle distribution.

\begin{figure}
	\includegraphics[scale=0.75]{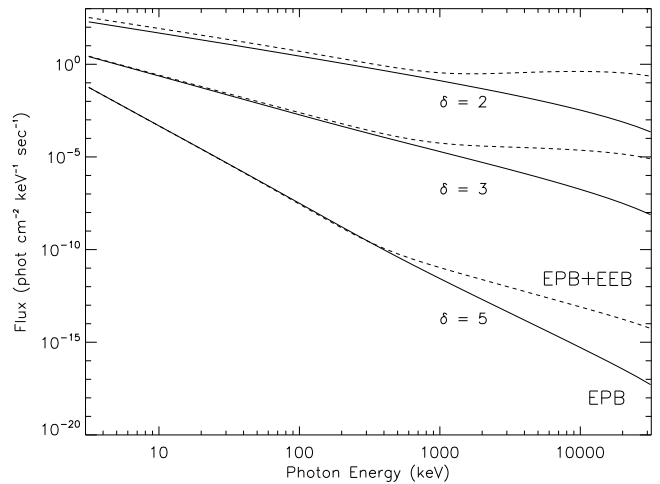}
	\caption{\label{fig:3} HXR spectra corresponding to power-law electron spectra with indices 2, 3, and 5 in the
		thick target model: total \mbox{\textit{e-i}}+\mbox{\textit{e-e}} flux (EPB+EEB, dashed) and \mbox{\textit{e-i}} flux only (EPB, solid)}
\end{figure}

\section{ELECTRON ENERGY SPECTRA IN 28.10.2003 AND 23.07.2002 FLARES}
For the gamma-rays, it is impossible to obtain high-quality spectra from local sources along the loop, but it can be established from the analysis of flare images (Fig. 2) that the integral spectrum in energy range above 100~keV corresponds to radiation only from the footpoints of the loop. Fig.~4 shows the HXR and gamma-ray spectra for these flares in the energy range from 150~keV up to 3~MeV.

\begin{figure*}
	\includegraphics[scale=0.38]{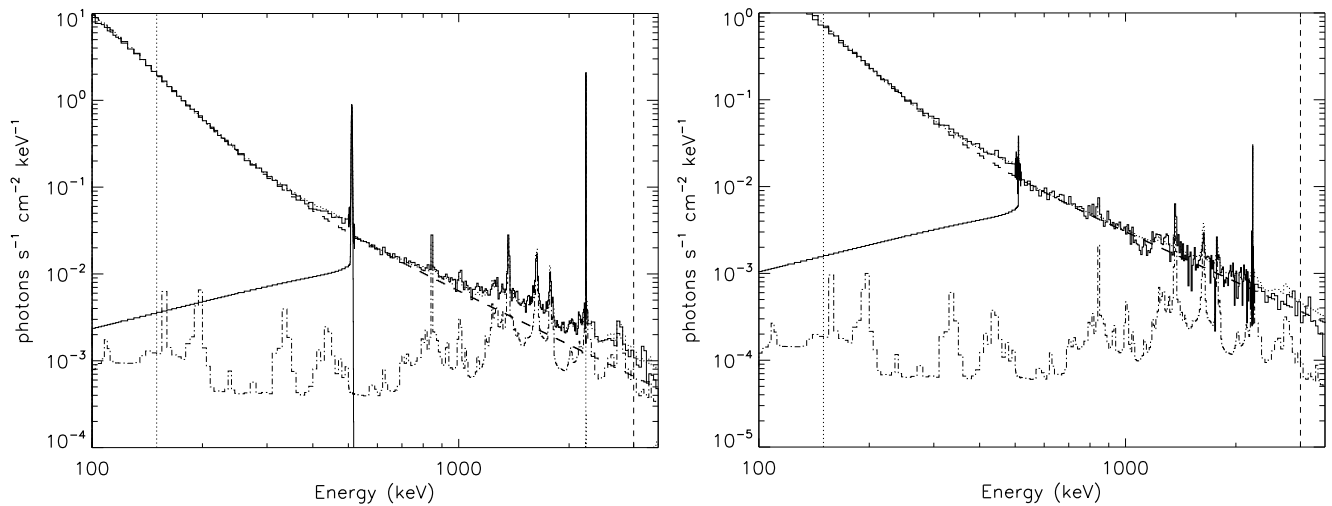}
	\caption{\label{fig:4} HXR and gamma-ray spectra of the October 28, 2003 flare in the time interval 11:08:40-
		11:09:00UT (left) and the July 27, 2002 flare in 00:28:40-00:29:00UT interval (right) according to
		the \textit{RHESSI} data (solid lines). The broken power-law bremsstrahlung spectra thick target model fits
		are shown including \mbox{\textit{e-e}} component (dashed lines), 511~keV and 2.2~MeV lines, and nuclear lines
		(lower curve) in the range from 150~keV to 3~MeV.
	}
\end{figure*}

For the 28.10.2003 flare, the bremsstrahlung HXR and gamma-ray spectrum parameters were determined after subtracting the radiation of 511~keV and 2.2~MeV lines, nuclear lines and the background gamma radiation. The bremsstrahlung spectrum is described by a broken power-law with the indices $\gamma_1=3.9$ and $\gamma_2=2.0$ and the break energy $\epsilon_{br} = 413$~keV. The recovered broken power-law electron spectrum in the thick-target model has a specific break at the energy $E_{br}= 1327$~keV, and spectral exponents $\delta_1=5.3$, $\delta_2=2.7$, in case \mbox{\textit{e-e}} bremsstrahlung is omitted. When approximating the HXR spectra, the contribution of \mbox{\textit{e-e}} radiation component results in a significantly softer electron spectrum above the break energy: $\delta_1=5.1$, $\delta_2=4.6$, and the break energy $E_{br}= 670$~keV. It is important to note that the HXR spectrum can be approximated using the formula (2) by a power-law electron spectrum with a spectral index $\delta_1=4.8$.

Similarly, for the flare of 23.07.2002, the following parameters of the broken power-law photon spectrum were obtained: $\gamma_=3.3,\gamma_2=2.1,\epsilon_{br}=412$~keV. We emphasize that for two flares similar spectral parameters and the same break energy in the photon spectrum occurred. Also, we point out that in the paper \cite{Emslie} spectral indices $\gamma$, obtained for the local sources in energy range from 50 to 120~keV and in the same time interval, vary from 2.9 to 3.2. The broken power-law electron spectrum obtained from the thick-target model without \mbox{\textit{e-e}} component as well as for the 28.10.2003 flare has a high-energy break at $E_{br} = 1274$~keV, spectral indices are $\delta_1=4.7$, $\delta_2=3.0$. Recovering of electron spectrum allowing for \mbox{\textit{e-e}} bremsstrahlung smoothes the break and leads to a drop in the high-energy region: spectral exponents are $\delta_1=4.4$, $\delta_2=5.1$, and $E_{br}= 1895$~keV. A power-law spectrum with the index $\delta=4.6$ was restored in the same way as for the flare of 28.10.2003.

We note for the same time interval in 23.07.2002 event, \cite{Hol03} paper report power-law electron energy spectra with index $\delta=4$, recovered from HXR spectrum in energy range up to 300~keV, which leads to an overestimation of the number of high energy electrons if \mbox{\textit{e-e}} bremsstrahlung is not taken into account. 

A slight difference in the spectrum parameters when approximated by a broken power-law leads to a significant change in the break energy and indicates the electron spectra in the considered energy range are power-law.

\section{CONCLUSIONS}

We performed the model calculations of \mbox{\textit{e-e}} and \mbox{\textit{e-i}} bremsstrahlung HXR flux from anisotropic electron distributions. Further we studied the electron energy spectra of two powerful solar flares SOL2003-10-28 T11:06:16 (\textit{GOES} class X17.2) and SOL2002-07-23 T00:18:16 (X4.8), analyzing HXR and gamma-ray spectra obtained from \textit{RHESSI} data.

The calculations revealed that the influence of \mbox{\textit{e-e}} bremsstrahlung on HXR flux noticeably depends on electron pitch angle distribution, flare viewing angle and hardness of electron energy spectra. Indeed, there is significant \mbox{\textit{e-e}} bremsstrahlung for isotropic pitch angle distribution, which can vary the photon spectral index $\gamma$ within 0.5 in case of hard electron energy spectra $\delta=2.5$ and low viewing angle $\alpha<10^\circ$. However, the growth of bremsstrahlung anisotropy with increasing electron energy requires allowing for the \mbox{\textit{e-i}} and \mbox{\textit{e-e}} bremsstrahlung cross-section angular dependence and the anisotropy of pitch-angular distributions of electrons. We show that for quasi-longitudinal pitch angle distribution and hard electron energy spectra the significant differences in photon spectral index ($\Delta\gamma>1$) with and without including \mbox{\textit{e-e}} bremsstrahlung can occur for photon energies $\epsilon\approx 100$~keV. Whereas \mbox{\textit{e-e}} bremsstrahlung does not notably affect the HXR spectra for quasi-transversal pitch electron distribution. The future study of high-energy electrons distribution anisotropy from observational data is required.

We allowed for \mbox{\textit{e-e}} bremsstrahlung approximating the HXR spectra of the 28 October 2003 and 23 July 2002 events. The contribution of \mbox{\textit{e-e}} radiation component results in a significantly softer electron spectrum above the break energy which leads to a more correct estimation of the number of high energy electrons. The inclusion of the \mbox{\textit{e-e}} component allows to restore a single power-law electron energy spectrum, while considering only the \mbox{\textit{e-i}} mechanism results in high-energy break $E_{br} > 1$~MeV. It is worth mentioning that for the 23 July 2002 flare there is a pronounced asymmetry towards the southern footpoint, starting from energies above 100~keV. The spatial separation of HXR sources at energies above 100~keV may be related to a feature of relativistic electron transport in flaring loops. The asymmetry of local gamma-ray sources points out the need for modeling the transport of accelerated electrons, considering the interdependence of the energy and angular parts of the electron distribution function.

\nocite{*}

\bibliography{apssamp}

\providecommand{\noopsort}[1]{}\providecommand{\singleletter}[1]{#1}%
\begin{thebibliography}{45}%
\makeatletter
\providecommand \@ifxundefined [1]{%
 \@ifx{#1\undefined}
}%
\providecommand \@ifnum [1]{%
 \ifnum #1\expandafter \@firstoftwo
 \else \expandafter \@secondoftwo
 \fi
}%
\providecommand \@ifx [1]{%
 \ifx #1\expandafter \@firstoftwo
 \else \expandafter \@secondoftwo
 \fi
}%
\providecommand \natexlab [1]{#1}%
\providecommand \enquote  [1]{``#1''}%
\providecommand \bibnamefont  [1]{#1}%
\providecommand \bibfnamefont [1]{#1}%
\providecommand \citenamefont [1]{#1}%
\providecommand \href@noop [0]{\@secondoftwo}%
\providecommand \href [0]{\begingroup \@sanitize@url \@href}%
\providecommand \@href[1]{\@@startlink{#1}\@@href}%
\providecommand \@@href[1]{\endgroup#1\@@endlink}%
\providecommand \@sanitize@url [0]{\catcode `\\12\catcode `\$12\catcode
  `\&12\catcode `\#12\catcode `\^12\catcode `\_12\catcode `\%12\relax}%
\providecommand \@@startlink[1]{}%
\providecommand \@@endlink[0]{}%
\providecommand \url  [0]{\begingroup\@sanitize@url \@url }%
\providecommand \@url [1]{\endgroup\@href {#1}{\urlprefix }}%
\providecommand \urlprefix  [0]{URL }%
\providecommand \Eprint [0]{\href }%
\providecommand \doibase [0]{https://doi.org/}%
\providecommand \selectlanguage [0]{\@gobble}%
\providecommand \bibinfo  [0]{\@secondoftwo}%
\providecommand \bibfield  [0]{\@secondoftwo}%
\providecommand \translation [1]{[#1]}%
\providecommand \BibitemOpen [0]{}%
\providecommand \bibitemStop [0]{}%
\providecommand \bibitemNoStop [0]{.\EOS\space}%
\providecommand \EOS [0]{\spacefactor3000\relax}%
\providecommand \BibitemShut  [1]{\csname bibitem#1\endcsname}%
\let\auto@bib@innerbib\@empty
\bibitem [{\citenamefont {Witten}(2001)}]{witten2001}%
  \BibitemOpen
  \bibfield  {author} {\bibinfo {author} {\bibfnamefont {E.}~\bibnamefont
  {Witten}},\ }\href@noop {} {} (\bibinfo {year} {2001}),\ \Eprint
  {https://arxiv.org/abs/hep-th/0106109} {hep-th/0106109} \BibitemShut
  {NoStop}%
\bibitem [{\citenamefont {Feynman}(1954)}]{feyn54}%
  \BibitemOpen
  \bibfield  {author} {\bibinfo {author} {\bibfnamefont {R.~P.}\ \bibnamefont
  {Feynman}},\ }\href@noop {} {\bibfield  {journal} {\bibinfo  {journal}
  {Phys.\ Rev.}\ }\textbf {\bibinfo {volume} {94}},\ \bibinfo {pages} {262}
  (\bibinfo {year} {1954})}\BibitemShut {NoStop}%
\bibitem [{\citenamefont {Einstein}\ \emph {et~al.}(1935)\citenamefont
  {Einstein}, \citenamefont {Podolsky},\ and\ \citenamefont {Rosen}}]{epr}%
  \BibitemOpen
  \bibfield  {author} {\bibinfo {author} {\bibfnamefont {A.}~\bibnamefont
  {Einstein}}, \bibinfo {author} {\bibfnamefont {{\relax Yu}.}~\bibnamefont
  {Podolsky}},\ and\ \bibinfo {author} {\bibfnamefont {N.}~\bibnamefont
  {Rosen}} (\bibinfo {collaboration} {EPR}),\ }\href@noop {} {\bibfield
  {journal} {\bibinfo  {journal} {Phys.\ Rev.}\ }\textbf {\bibinfo {volume}
  {47}},\ \bibinfo {pages} {777} (\bibinfo {year} {1935})}\BibitemShut
  {NoStop}%
\bibitem [{\citenamefont {Beutler}(1994{\natexlab{a}})}]{Beutler1994}%
  \BibitemOpen
  \bibfield  {author} {\bibinfo {author} {\bibfnamefont {E.}~\bibnamefont
  {Beutler}},\ }in\ \href@noop {} {\emph {\bibinfo {booktitle} {Williams
  Hematology}}},\ Vol.~\bibinfo {volume} {2},\ \bibinfo {editor} {edited by\
  \bibinfo {editor} {\bibfnamefont {E.}~\bibnamefont {Beutler}}, \bibinfo
  {editor} {\bibfnamefont {M.~A.}\ \bibnamefont {Lichtman}}, \bibinfo {editor}
  {\bibfnamefont {B.~W.}\ \bibnamefont {Coller}},\ and\ \bibinfo {editor}
  {\bibfnamefont {T.~S.}\ \bibnamefont {Kipps}}}\ (\bibinfo  {publisher}
  {McGraw-Hill},\ \bibinfo {address} {New York},\ \bibinfo {year} {1994})\
  Chap.~\bibinfo {chapter} {7}, pp.\ \bibinfo {pages} {654--662},\ \bibinfo
  {edition} {5th}\ ed.\BibitemShut {Stop}%
\bibitem [{\citenamefont {Birell}\ and\ \citenamefont {Davies}(1982)}]{Bire82}%
  \BibitemOpen
  \bibfield  {author} {\bibinfo {author} {\bibfnamefont {N.~D.}\ \bibnamefont
  {Birell}}\ and\ \bibinfo {author} {\bibfnamefont {P.~C.~W.}\ \bibnamefont
  {Davies}},\ }\href@noop {} {\emph {\bibinfo {title} {Quantum Fields in Curved
  Space}}}\ (\bibinfo  {publisher} {Cambridge University Press},\ \bibinfo
  {year} {1982})\BibitemShut {NoStop}%
\bibitem [{\citenamefont {G.~P.~Berman}\ and\ \citenamefont
  {F.~M.~Izrailev}(1983)}]{Berman1983}%
  \BibitemOpen
  \bibfield  {author} {\bibinfo {author} {\bibfnamefont {J.}~\bibnamefont
  {G.~P.~Berman}}\ and\ \bibinfo {author} {\bibfnamefont {J.}~\bibnamefont
  {F.~M.~Izrailev}},\ }\bibfield  {title} {\bibinfo {title} {Stability of
  nonlinear modes},\ }\href@noop {} {\bibfield  {journal} {\bibinfo  {journal}
  {Physica D}\ }\textbf {\bibinfo {volume} {88}},\ \bibinfo {pages} {445}
  (\bibinfo {year} {1983})}\BibitemShut {NoStop}%
\bibitem [{\citenamefont {Davies}\ and\ \citenamefont
  {Parns}(1988)}]{Davies1998}%
  \BibitemOpen
  \bibfield  {author} {\bibinfo {author} {\bibfnamefont {E.~B.}\ \bibnamefont
  {Davies}}\ and\ \bibinfo {author} {\bibfnamefont {L.}~\bibnamefont {Parns}},\
  }\bibfield  {title} {\bibinfo {title} {Trapped modes in acoustic
  waveguides},\ }\href@noop {} {\bibfield  {journal} {\bibinfo  {journal} {Q.
  J. Mech. Appl. Math.}\ }\textbf {\bibinfo {volume} {51}},\ \bibinfo {pages}
  {477} (\bibinfo {year} {1988})}\BibitemShut {NoStop}%
\bibitem [{Note1()}]{Note1}%
  \BibitemOpen
  \bibinfo {note} {Automatically placing footnotes into the bibliography
  requires using BibTeX to compile the bibliography.}\BibitemShut {Stop}%
\bibitem [{\citenamefont {Beutler}(1994{\natexlab{b}})}]{Beutler1994a}%
  \BibitemOpen
  \bibfield  {author} {\bibinfo {author} {\bibfnamefont {E.}~\bibnamefont
  {Beutler}},\ }in\ \href@noop {} {\emph {\bibinfo {booktitle} {Williams
  Hematology}}},\ Vol.~\bibinfo {volume} {2},\ \bibinfo {editor} {edited by\
  \bibinfo {editor} {\bibfnamefont {E.}~\bibnamefont {Beutler}}, \bibinfo
  {editor} {\bibfnamefont {M.~A.}\ \bibnamefont {Lichtman}}, \bibinfo {editor}
  {\bibfnamefont {B.~W.}\ \bibnamefont {Coller}},\ and\ \bibinfo {editor}
  {\bibfnamefont {T.~S.}\ \bibnamefont {Kipps}}}\ (\bibinfo  {publisher}
  {McGraw-Hill},\ \bibinfo {address} {New York},\ \bibinfo {year} {1994})\
  \bibinfo {edition} {5th}\ ed.,\ Chap.~\bibinfo {chapter} {7}, pp.\ \bibinfo
  {pages} {654--662}\BibitemShut {NoStop}%
\bibitem [{\citenamefont {Knuth}(1973)}]{inbook-full}%
  \BibitemOpen
  \bibfield  {author} {\bibinfo {author} {\bibfnamefont {D.~E.}\ \bibnamefont
  {Knuth}},\ }in\ \href@noop {} {\emph {\bibinfo {booktitle} {Fundamental
  Algorithms}}},\ \bibinfo {series} {The Art of Computer Programming},
  Vol.~\bibinfo {volume} {1}\ (\bibinfo  {publisher} {Addison-Wesley},\
  \bibinfo {address} {Reading, Massachusetts},\ \bibinfo {year}
  {\noopsort{1973b}1973})\ \bibinfo {type} {Section}\ \bibinfo {chapter} {1.2},
  pp.\ \bibinfo {pages} {10--119},\ \bibinfo {edition} {2nd}\ ed.,\ \bibinfo
  {note} {a full INBOOK entry}\BibitemShut {NoStop}%
\bibitem [{\citenamefont {Smith}\ and\ \citenamefont
  {Johnson}(2005)}]{Smith2005}%
  \BibitemOpen
  \bibfield  {author} {\bibinfo {author} {\bibfnamefont {J.~S.}\ \bibnamefont
  {Smith}}\ and\ \bibinfo {author} {\bibfnamefont {G.~W.}\ \bibnamefont
  {Johnson}},\ }\href@noop {} {\bibfield  {journal} {\bibinfo  {journal}
  {Philos. Trans. R. Soc. London, Ser. B}\ }\textbf {\bibinfo {volume} {777}},\
  \bibinfo {pages} {1395} (\bibinfo {year} {2005})}\BibitemShut {NoStop}%
\bibitem [{\citenamefont {Smith}\ \emph
  {et~al.}(2010{\natexlab{a}})\citenamefont {Smith}, \citenamefont {Johnson},\
  and\ \citenamefont {Miller}}]{Smith2010}%
  \BibitemOpen
  \bibfield  {author} {\bibinfo {author} {\bibfnamefont {W.~J.}\ \bibnamefont
  {Smith}}, \bibinfo {author} {\bibfnamefont {T.~J.}\ \bibnamefont {Johnson}},\
  and\ \bibinfo {author} {\bibfnamefont {B.~G.}\ \bibnamefont {Miller}},\
  }\bibfield  {title} {\bibinfo {title} {Surface chemistry and preferential
  crystal orientation on a silicon surface}} (\bibinfo {year}
  {2010}{\natexlab{a}}),\ \bibinfo {note} {{J. Appl. Phys.}
  (unpublished)}\BibitemShut {NoStop}%
\bibitem [{\citenamefont {Smith}\ \emph
  {et~al.}(2010{\natexlab{b}})\citenamefont {Smith}, \citenamefont {Johnson},\
  and\ \citenamefont {Klein}}]{Smith2010a}%
  \BibitemOpen
  \bibfield  {author} {\bibinfo {author} {\bibfnamefont {V.~K.}\ \bibnamefont
  {Smith}}, \bibinfo {author} {\bibfnamefont {K.}~\bibnamefont {Johnson}},\
  and\ \bibinfo {author} {\bibfnamefont {M.~O.}\ \bibnamefont {Klein}},\
  }\bibfield  {title} {\bibinfo {title} {Surface chemistry and preferential
  crystal orientation on a silicon surface}} (\bibinfo {year}
  {2010}{\natexlab{b}}),\ \bibinfo {note} {{J. Appl. Phys.}
  (submitted)}\BibitemShut {NoStop}%
\bibitem [{\citenamefont {{\"{U}}nderwood}\ \emph {et~al.}(1988)\citenamefont
  {{\"{U}}nderwood}, \citenamefont {{\~N}et},\ and\ \citenamefont
  {{\={P}}ot}}]{unpublished-full}%
  \BibitemOpen
  \bibfield  {author} {\bibinfo {author} {\bibfnamefont {U.}~\bibnamefont
  {{\"{U}}nderwood}}, \bibinfo {author} {\bibfnamefont {N.}~\bibnamefont
  {{\~N}et}},\ and\ \bibinfo {author} {\bibfnamefont {P.}~\bibnamefont
  {{\={P}}ot}},\ }\bibfield  {title} {\bibinfo {title} {Lower bounds for
  wishful research results}} (\bibinfo {year} {1988}),\ \bibinfo {note} {talk
  at Fanstord University (A full UNPUBLISHED entry)}\BibitemShut {NoStop}%
\bibitem [{\citenamefont {Johnson}\ \emph {et~al.}(2007)\citenamefont
  {Johnson}, \citenamefont {Miller},\ and\ \citenamefont
  {Smith}}]{JohnsonMillerSmith2007}%
  \BibitemOpen
  \bibfield  {author} {\bibinfo {author} {\bibfnamefont {M.~P.}\ \bibnamefont
  {Johnson}}, \bibinfo {author} {\bibfnamefont {K.~L.}\ \bibnamefont
  {Miller}},\ and\ \bibinfo {author} {\bibfnamefont {K.}~\bibnamefont
  {Smith}},\ }\href@noop {} {}\bibinfo {howpublished} {personal communication}
  (\bibinfo {year} {2007})\BibitemShut {NoStop}%
\bibitem [{\citenamefont {Smith}(2007{\natexlab{a}})}]{Smith2007}%
  \BibitemOpen
  \bibinfo {editor} {\bibfnamefont {J.}~\bibnamefont {Smith}},\ ed.,\
  \href@noop {} {\emph {\bibinfo {title} {AIP Conf. Proc.}}},\ Vol.\ \bibinfo
  {volume} {841}\ (\bibinfo {year} {2007})\BibitemShut {NoStop}%
\bibitem [{\citenamefont {Oz}\ and\ \citenamefont
  {Yannakakis}(1983)}]{proceedings-full}%
  \BibitemOpen
  \bibinfo {editor} {\bibfnamefont {W.~V.}\ \bibnamefont {Oz}}\ and\ \bibinfo
  {editor} {\bibfnamefont {M.}~\bibnamefont {Yannakakis}},\ eds.,\ \href@noop
  {} {\emph {\bibinfo {title} {Proc. Fifteenth Annual}}},\ \bibinfo {series}
  {All ACM Conferences}\ No.~\bibinfo {number} {17},\ \bibinfo {organization}
  {ACM}\ (\bibinfo  {publisher} {Academic Press},\ \bibinfo {address}
  {Boston},\ \bibinfo {year} {1983})\ \bibinfo {note} {a full PROCEEDINGS
  entry}\BibitemShut {NoStop}%
\bibitem [{\citenamefont {Burstyn}(2004)}]{Burstyn2004}%
  \BibitemOpen
  \bibfield  {author} {\bibinfo {author} {\bibfnamefont {Y.}~\bibnamefont
  {Burstyn}},\ }\bibfield  {title} {\bibinfo {title} {{Proceedings of the 5th
  International Molecular Beam Epitaxy Conference, Santa Fe, NM}}} (\bibinfo
  {year} {2004}),\ \bibinfo {note} {(unpublished)}\BibitemShut {NoStop}%
\bibitem [{\citenamefont {Quinn}(2001)}]{Quinn2001}%
  \BibitemOpen
  \bibinfo {editor} {\bibfnamefont {B.}~\bibnamefont {Quinn}},\ ed.,\
  \href@noop {} {\emph {\bibinfo {title} {{Proceedings of the 2003 Particle
  Accelerator Conference, Portland, OR, 12-16 May 2005}}}}\ (\bibinfo
  {publisher} {Wiley},\ \bibinfo {address} {New York},\ \bibinfo {year}
  {2001})\ \bibinfo {note} {albeit the conference was held in 2005, it was the
  2003 conference, and the proceedings were published in 2001; go
  figure}\BibitemShut {NoStop}%
\bibitem [{\citenamefont {Agarwal}(2001)}]{Agarwal2001}%
  \BibitemOpen
  \bibfield  {author} {\bibinfo {author} {\bibfnamefont {A.~G.}\ \bibnamefont
  {Agarwal}},\ }\bibfield  {title} {\bibinfo {title} {{Proceedings of the Fifth
  Low Temperature Conference, Madison, WI, 1999}},\ }\href@noop {} {\bibfield
  {journal} {\bibinfo  {journal} {Semiconductors}\ }\textbf {\bibinfo {volume}
  {66}},\ \bibinfo {pages} {1238} (\bibinfo {year} {2001})}\BibitemShut
  {NoStop}%
\bibitem [{\citenamefont {Smith}(2001)}]{SmithDA01}%
  \BibitemOpen
  \bibfield  {author} {\bibinfo {author} {\bibfnamefont {R.}~\bibnamefont
  {Smith}},\ }\bibfield  {title} {\bibinfo {title} {Hummingbirds are our
  friends},\ }\href@noop {} {\bibfield  {journal} {\bibinfo  {journal} {J.
  Appl. Phys. (these proceedings)}\ } (\bibinfo {year} {2001})},\ \bibinfo
  {note} {abstract No. DA-01}\BibitemShut {NoStop}%
\bibitem [{\citenamefont {Smith}(2007{\natexlab{b}})}]{Smith2007a}%
  \BibitemOpen
  \bibfield  {author} {\bibinfo {author} {\bibfnamefont {J.}~\bibnamefont
  {Smith}},\ }\href@noop {} {\bibfield  {journal} {\bibinfo  {journal} {Proc.
  SPIE}\ }\textbf {\bibinfo {volume} {124}},\ \bibinfo {pages} {367} (\bibinfo
  {year} {2007}{\natexlab{b}})},\ \bibinfo {note} {required title is
  missing}\BibitemShut {NoStop}%
\bibitem [{\citenamefont {T{\'{e}}rrific}(1988)}]{techreport-full}%
  \BibitemOpen
  \bibfield  {author} {\bibinfo {author} {\bibfnamefont {T.}~\bibnamefont
  {T{\'{e}}rrific}},\ }\href@noop {} {\emph {\bibinfo {title} {An {$O(n \log n
  / \! \log\log n)$} Sorting Algorithm}}},\ \bibinfo {type} {Wishful Research
  Result}\ \bibinfo {number} {7}\ (\bibinfo  {institution} {Fanstord
  University},\ \bibinfo {address} {Computer Science Department, Fanstord,
  California},\ \bibinfo {year} {1988})\ \bibinfo {note} {a full TECHREPORT
  entry}\BibitemShut {NoStop}%
\bibitem [{\citenamefont {Nelson}(1999{\natexlab{a}})}]{Nelson1999}%
  \BibitemOpen
  \bibfield  {author} {\bibinfo {author} {\bibfnamefont {J.}~\bibnamefont
  {Nelson}},\ }\href@noop {} {}\bibinfo {type} {{TWI Report}}\ \bibinfo
  {number} {666/1999}\ (\bibinfo {year} {Jan.~1999})\ \bibinfo {note} {required
  institution missing}\BibitemShut {NoStop}%
\bibitem [{\citenamefont {Fields}(2005)}]{Fields2005}%
  \BibitemOpen
  \bibfield  {author} {\bibinfo {author} {\bibfnamefont {W.~K.}\ \bibnamefont
  {Fields}},\ }\href@noop {} {}\bibinfo {type} {{ECE Report No.}}\ \bibinfo
  {number} {AL944}\ (\bibinfo {year} {2005})\ \bibinfo {note} {required
  institution missing}\BibitemShut {NoStop}%
\bibitem [{\citenamefont {Zalkins}(2008)}]{Zalkins2008}%
  \BibitemOpen
  \bibfield  {author} {\bibinfo {author} {\bibfnamefont {Y.~M.}\ \bibnamefont
  {Zalkins}},\ }\href@noop {} {}\bibinfo {howpublished} {e-print
  arXiv:cond-mat/040426} (\bibinfo {year} {2008})\BibitemShut {NoStop}%
\bibitem [{\citenamefont {Nelson}(2005)}]{Nelson2005}%
  \BibitemOpen
  \bibfield  {author} {\bibinfo {author} {\bibfnamefont {J.}~\bibnamefont
  {Nelson}},\ }\href@noop {} {}\bibinfo {howpublished} {{U.S. Patent No.}
  5,693,000} (\bibinfo {year} {12~Dec.~2005})\BibitemShut {NoStop}%
\bibitem [{\citenamefont {Nelson}(1999{\natexlab{b}})}]{Nelson1999a}%
  \BibitemOpen
  \bibfield  {author} {\bibinfo {author} {\bibfnamefont {J.~K.}\ \bibnamefont
  {Nelson}},\ }\href@noop {} {\bibinfo {type} {M.{S}. thesis}},\ \bibinfo
  {school} {New York University} (\bibinfo {year}
  {1999}{\natexlab{b}})\BibitemShut {NoStop}%
\bibitem [{\citenamefont {Masterly}(1988)}]{mastersthesis-full}%
  \BibitemOpen
  \bibfield  {author} {\bibinfo {author} {\bibfnamefont {{\'{E}}.}~\bibnamefont
  {Masterly}},\ }\emph {\bibinfo {title} {Mastering Thesis Writing}},\
  \href@noop {} {\bibinfo {type} {Master's project}},\ \bibinfo  {school}
  {Stanford University}, \bibinfo {address} {English Department} (\bibinfo
  {year} {1988}),\ \bibinfo {note} {a full MASTERSTHESIS entry}\BibitemShut
  {NoStop}%
\bibitem [{\citenamefont {Smith}(2003)}]{Smith2003}%
  \BibitemOpen
  \bibfield  {author} {\bibinfo {author} {\bibfnamefont {S.~M.}\ \bibnamefont
  {Smith}},\ }\href@noop {} {\bibinfo {type} {{Ph.D.} thesis}},\ \bibinfo
  {school} {Massachusetts Institute of Technology} (\bibinfo {year}
  {2003})\BibitemShut {NoStop}%
\bibitem [{\citenamefont {Kawa}\ and\ \citenamefont {Lin}(2003)}]{KawaLin2003}%
  \BibitemOpen
  \bibfield  {author} {\bibinfo {author} {\bibfnamefont {S.~R.}\ \bibnamefont
  {Kawa}}\ and\ \bibinfo {author} {\bibfnamefont {S.-J.}\ \bibnamefont {Lin}},\
  }\href@noop {} {\bibfield  {journal} {\bibinfo  {journal} {J. Geophys. Res.}\
  }\textbf {\bibinfo {volume} {108}},\ \bibinfo {pages} {4201} (\bibinfo {year}
  {2003})},\ \bibinfo {note} {{DOI:10.1029/2002JD002268}}\BibitemShut {NoStop}%
\bibitem [{\citenamefont {Phony-Baloney}(1988)}]{phdthesis-full}%
  \BibitemOpen
  \bibfield  {author} {\bibinfo {author} {\bibfnamefont {F.~P.}\ \bibnamefont
  {Phony-Baloney}},\ }\emph {\bibinfo {title} {Fighting Fire with Fire:
  Festooning {F}rench Phrases}},\ \href@noop {} {\bibinfo {type} {{PhD}
  dissertation}},\ \bibinfo  {school} {Fanstord University}, \bibinfo {address}
  {Department of French} (\bibinfo {year} {1988}),\ \bibinfo {note} {a full
  PHDTHESIS entry}\BibitemShut {NoStop}%
\bibitem [{\citenamefont {Knuth}(1981)}]{book-full}%
  \BibitemOpen
  \bibfield  {author} {\bibinfo {author} {\bibfnamefont {D.~E.}\ \bibnamefont
  {Knuth}},\ }\href@noop {} {\emph {\bibinfo {title} {Seminumerical
  Algorithms}}},\ \bibinfo {edition} {2nd}\ ed.,\ \bibinfo {series} {The Art of
  Computer Programming}, Vol.~\bibinfo {volume} {2}\ (\bibinfo  {publisher}
  {Addison-Wesley},\ \bibinfo {address} {Reading, Massachusetts},\ \bibinfo
  {year} {\noopsort{1973c}1981})\ \bibinfo {note} {a full BOOK
  entry}\BibitemShut {NoStop}%
\bibitem [{\citenamefont {Knvth}(1988)}]{booklet-full}%
  \BibitemOpen
  \bibfield  {author} {\bibinfo {author} {\bibfnamefont {J.~C.}\ \bibnamefont
  {Knvth}},\ }\href@noop {} {\bibinfo {title} {The programming of computer
  art}},\ \bibinfo {howpublished} {Vernier Art Center},\ \bibinfo {address}
  {Stanford, California} (\bibinfo {year} {1988}),\ \bibinfo {note} {a full
  BOOKLET entry}\BibitemShut {NoStop}%
\bibitem [{\citenamefont {Ballagh}\ and\ \citenamefont
  {Savage}(2000{\natexlab{a}})}]{ballagh2000}%
  \BibitemOpen
  \bibfield  {author} {\bibinfo {author} {\bibfnamefont {R.}~\bibnamefont
  {Ballagh}}\ and\ \bibinfo {author} {\bibfnamefont {C.}~\bibnamefont
  {Savage}},\ }\bibinfo {title} {Bose-einstein condensation: from atomic
  physics to quantum fluids, proceedings of the 13th physics summer school}\
  (\bibinfo  {publisher} {World Scientific},\ \bibinfo {address} {Singapore},\
  \bibinfo {year} {2000})\ \Eprint {https://arxiv.org/abs/cond-mat/0008070}
  {cond-mat/0008070} \BibitemShut {NoStop}%
\bibitem [{\citenamefont {Ballagh}\ and\ \citenamefont
  {Savage}(2000{\natexlab{b}})}]{ballagh2000a}%
  \BibitemOpen
  \bibfield  {author} {\bibinfo {author} {\bibfnamefont {R.}~\bibnamefont
  {Ballagh}}\ and\ \bibinfo {author} {\bibfnamefont {C.}~\bibnamefont
  {Savage}},\ }\bibfield  {title} {\bibinfo {title} {Bose-einstein
  condensation: from atomic physics to quantum fluids},\ }in\ \href@noop {}
  {\emph {\bibinfo {booktitle} {Proceedings of the 13th Physics Summer
  School}}},\ \bibinfo {editor} {edited by\ \bibinfo {editor} {\bibfnamefont
  {C.}~\bibnamefont {Savage}}\ and\ \bibinfo {editor} {\bibfnamefont
  {M.}~\bibnamefont {Das}}}\ (\bibinfo  {publisher} {World Scientific},\
  \bibinfo {address} {Singapore},\ \bibinfo {year} {2000})\ \Eprint
  {https://arxiv.org/abs/cond-mat/0008070} {cond-mat/0008070} \BibitemShut
  {NoStop}%
\bibitem [{\citenamefont {Opechowski}\ and\ \citenamefont
  {Guccione}(1965{\natexlab{a}})}]{Magnetism}%
  \BibitemOpen
  \bibfield  {author} {\bibinfo {author} {\bibfnamefont {W.}~\bibnamefont
  {Opechowski}}\ and\ \bibinfo {author} {\bibfnamefont {R.}~\bibnamefont
  {Guccione}},\ }\bibinfo {title} {Introduction to the theory of normal
  metals},\ in\ \href@noop {} {\emph {\bibinfo {booktitle} {Magnetism}}},\
  Vol.\ \bibinfo {volume} {IIa},\ \bibinfo {editor} {edited by\ \bibinfo
  {editor} {\bibfnamefont {G.~T.}\ \bibnamefont {Rado}}\ and\ \bibinfo {editor}
  {\bibfnamefont {H.}~\bibnamefont {Suhl}}}\ (\bibinfo  {publisher} {Academic
  Press},\ \bibinfo {address} {New York},\ \bibinfo {year} {1965})\ p.\
  \bibinfo {pages} {105}\BibitemShut {NoStop}%
\bibitem [{\citenamefont {Opechowski}\ and\ \citenamefont
  {Guccione}(1965{\natexlab{b}})}]{Magnetisma}%
  \BibitemOpen
  \bibfield  {author} {\bibinfo {author} {\bibfnamefont {W.}~\bibnamefont
  {Opechowski}}\ and\ \bibinfo {author} {\bibfnamefont {R.}~\bibnamefont
  {Guccione}},\ }\bibfield  {title} {\bibinfo {title} {Introduction to the
  theory of normal metals},\ }in\ \href@noop {} {\emph {\bibinfo {booktitle}
  {Magnetism}}},\ Vol.\ \bibinfo {volume} {IIa},\ \bibinfo {editor} {edited by\
  \bibinfo {editor} {\bibfnamefont {G.~T.}\ \bibnamefont {Rado}}\ and\ \bibinfo
  {editor} {\bibfnamefont {H.}~\bibnamefont {Suhl}}}\ (\bibinfo  {publisher}
  {Academic Press},\ \bibinfo {address} {New York},\ \bibinfo {year} {1965})\
  p.\ \bibinfo {pages} {105}\BibitemShut {NoStop}%
\bibitem [{\citenamefont {Opechowski}\ and\ \citenamefont
  {Guccione}(1965{\natexlab{c}})}]{Magnetismb}%
  \BibitemOpen
  \bibfield  {author} {\bibinfo {author} {\bibfnamefont {W.}~\bibnamefont
  {Opechowski}}\ and\ \bibinfo {author} {\bibfnamefont {R.}~\bibnamefont
  {Guccione}},\ }\bibfield  {title} {\bibinfo {title} {Introduction to the
  theory of normal metals},\ }in\ \href@noop {} {\emph {\bibinfo {booktitle}
  {Magnetism}}},\ Vol.\ \bibinfo {volume} {IIa},\ \bibinfo {editor} {edited by\
  \bibinfo {editor} {\bibfnamefont {G.~T.}\ \bibnamefont {Rado}}\ and\ \bibinfo
  {editor} {\bibfnamefont {H.}~\bibnamefont {Suhl}}}\ (\bibinfo  {publisher}
  {Academic Press},\ \bibinfo {address} {New York},\ \bibinfo {year} {1965})\
  p.\ \bibinfo {pages} {105}\BibitemShut {NoStop}%
\bibitem [{\citenamefont {Smith}(1980{\natexlab{a}})}]{Smith80}%
  \BibitemOpen
  \bibfield  {author} {\bibinfo {author} {\bibfnamefont {J.~M.}\ \bibnamefont
  {Smith}},\ }\bibinfo {title} {Molecular dynamics}\ (\bibinfo  {publisher}
  {Academic},\ \bibinfo {address} {New York},\ \bibinfo {year}
  {1980})\BibitemShut {NoStop}%
\bibitem [{\citenamefont {Zakharov}\ and\ \citenamefont {Shabat}(1971)}]{ZS71}%
  \BibitemOpen
  \bibfield  {author} {\bibinfo {author} {\bibfnamefont {V.~E.}\ \bibnamefont
  {Zakharov}}\ and\ \bibinfo {author} {\bibfnamefont {A.~B.}\ \bibnamefont
  {Shabat}},\ }\bibfield  {title} {\bibinfo {title} {Exact theory of
  two-dimensional self-focusing and one-dimensional self-modulation of waves in
  nonlinear media},\ }\href@noop {} {\bibfield  {journal} {\bibinfo  {journal}
  {Zh. Eksp. Teor. Fiz.}\ }\textbf {\bibinfo {volume} {61}},\ \bibinfo {pages}
  {118} (\bibinfo {year} {1971})},\ \translation{Sov. Phys. JETP \textbf{34},
  62 (1972)}\BibitemShut {NoStop}%
\bibitem [{\citenamefont {Smith}(1980{\natexlab{b}})}]{Smith80a}%
  \BibitemOpen
  \bibfield  {author} {\bibinfo {author} {\bibfnamefont {J.~M.}\ \bibnamefont
  {Smith}},\ }in\ \href@noop {} {\emph {\bibinfo {booktitle} {Molecular
  Dynamics}}},\ \bibinfo {editor} {edited by\ \bibinfo {editor} {\bibfnamefont
  {C.}~\bibnamefont {Brown}}}\ (\bibinfo  {publisher} {Academic},\ \bibinfo
  {address} {New York},\ \bibinfo {year} {1980})\BibitemShut {NoStop}%
\bibitem [{\citenamefont {Lincoll}(1977)}]{incollection-full}%
  \BibitemOpen
  \bibfield  {author} {\bibinfo {author} {\bibfnamefont {D.~D.}\ \bibnamefont
  {Lincoll}},\ }\bibfield  {title} {\bibinfo {title} {Semigroups of
  recurrences},\ }in\ \href@noop {} {\emph {\bibinfo {booktitle} {High Speed
  Computer and Algorithm Organization}}},\ \bibinfo {series and number}
  {\bibinfo {series} {Fast Computers}\ No.~\bibinfo {number} {23}},\ \bibinfo
  {editor} {edited by\ \bibinfo {editor} {\bibfnamefont {D.~J.}\ \bibnamefont
  {Lipcoll}}, \bibinfo {editor} {\bibfnamefont {D.~H.}\ \bibnamefont
  {Lawrie}},\ and\ \bibinfo {editor} {\bibfnamefont {A.~H.}\ \bibnamefont
  {Sameh}}}\ (\bibinfo  {publisher} {Academic Press},\ \bibinfo {address} {New
  York},\ \bibinfo {year} {1977})\ \bibinfo {edition} {3rd}\ ed.,\ \bibinfo
  {type} {Part}~\bibinfo {chapter} {3}, pp.\ \bibinfo {pages} {179--183},\
  \bibinfo {note} {a full INCOLLECTION entry}\BibitemShut {NoStop}%
\bibitem [{\citenamefont {Oaho}\ \emph {et~al.}(1983)\citenamefont {Oaho},
  \citenamefont {Ullman},\ and\ \citenamefont
  {Yannakakis}}]{inproceedings-full}%
  \BibitemOpen
  \bibfield  {author} {\bibinfo {author} {\bibfnamefont {A.~V.}\ \bibnamefont
  {Oaho}}, \bibinfo {author} {\bibfnamefont {J.~D.}\ \bibnamefont {Ullman}},\
  and\ \bibinfo {author} {\bibfnamefont {M.}~\bibnamefont {Yannakakis}},\
  }\bibfield  {title} {\bibinfo {title} {On notions of information transfer in
  {VLSI} circuits},\ }in\ \href@noop {} {\emph {\bibinfo {booktitle} {Proc.
  Fifteenth Annual ACM}}},\ \bibinfo {address} {Boston, 1982},\ \bibinfo
  {series and number} {\bibinfo {series} {All ACM Conferences}\ No.~\bibinfo
  {number} {17}},\ \bibinfo {editor} {edited by\ \bibinfo {editor}
  {\bibfnamefont {W.~V.}\ \bibnamefont {Oz}}\ and\ \bibinfo {editor}
  {\bibfnamefont {M.}~\bibnamefont {Yannakakis}}},\ \bibinfo {organization}
  {ACM}\ (\bibinfo  {publisher} {Academic Press},\ \bibinfo {address} {New
  York},\ \bibinfo {year} {1983})\ pp.\ \bibinfo {pages} {133--139},\ \bibinfo
  {note} {a full INPROCEDINGS entry}\BibitemShut {NoStop}%
\bibitem [{\citenamefont {Manmaker}(1986)}]{manual-full}%
  \BibitemOpen
  \bibfield  {author} {\bibinfo {author} {\bibfnamefont {L.}~\bibnamefont
  {Manmaker}},\ }\href@noop {} {\emph {\bibinfo {title} {The Definitive
  Computer Manual}}},\ \bibinfo {organization} {Chips-R-Us},\ \bibinfo
  {address} {Silicon Valley},\ \bibinfo {edition} {silver}\ ed. (\bibinfo
  {year} {1986}),\ \bibinfo {note} {a full MANUAL entry}\BibitemShut {NoStop}%
\end{thebibliography}%


\providecommand{\noopsort}[1]{}\providecommand{\singleletter}[1]{#1}%
\begin{thebibliography}{19}%
\makeatletter
\providecommand \@ifxundefined [1]{%
 \@ifx{#1\undefined}
}%
\providecommand \@ifnum [1]{%
 \ifnum #1\expandafter \@firstoftwo
 \else \expandafter \@secondoftwo
 \fi
}%
\providecommand \@ifx [1]{%
 \ifx #1\expandafter \@firstoftwo
 \else \expandafter \@secondoftwo
 \fi
}%
\providecommand \natexlab [1]{#1}%
\providecommand \enquote  [1]{``#1''}%
\providecommand \bibnamefont  [1]{#1}%
\providecommand \bibfnamefont [1]{#1}%
\providecommand \citenamefont [1]{#1}%
\providecommand \href@noop [0]{\@secondoftwo}%
\providecommand \href [0]{\begingroup \@sanitize@url \@href}%
\providecommand \@href[1]{\@@startlink{#1}\@@href}%
\providecommand \@@href[1]{\endgroup#1\@@endlink}%
\providecommand \@sanitize@url [0]{\catcode `\\12\catcode `\$12\catcode
  `\&12\catcode `\#12\catcode `\^12\catcode `\_12\catcode `\%12\relax}%
\providecommand \@@startlink[1]{}%
\providecommand \@@endlink[0]{}%
\providecommand \url  [0]{\begingroup\@sanitize@url \@url }%
\providecommand \@url [1]{\endgroup\@href {#1}{\urlprefix }}%
\providecommand \urlprefix  [0]{URL }%
\providecommand \Eprint [0]{\href }%
\providecommand \doibase [0]{https://doi.org/}%
\providecommand \selectlanguage [0]{\@gobble}%
\providecommand \bibinfo  [0]{\@secondoftwo}%
\providecommand \bibfield  [0]{\@secondoftwo}%
\providecommand \translation [1]{[#1]}%
\providecommand \BibitemOpen [0]{}%
\providecommand \bibitemStop [0]{}%
\providecommand \bibitemNoStop [0]{.\EOS\space}%
\providecommand \EOS [0]{\spacefactor3000\relax}%
\providecommand \BibitemShut  [1]{\csname bibitem#1\endcsname}%
\let\auto@bib@innerbib\@empty
\bibitem [{\citenamefont {Aschwanden}(2004)}]{Asch}%
  \BibitemOpen
  \bibfield  {author} {\bibinfo {author} {\bibfnamefont {M.~J.}\ \bibnamefont
  {Aschwanden}},\ }\href@noop {} {\emph {\bibinfo {title} {Physics of the Solar
  Corona}}}\ (\bibinfo  {publisher} {Springer-Verlag Berlin Heidelberg},\
  \bibinfo {year} {2004})\BibitemShut {NoStop}%
\bibitem [{\citenamefont {Zheleznyakov}(1977)}]{Zhel}%
  \BibitemOpen
  \bibfield  {author} {\bibinfo {author} {\bibfnamefont {V.~V.}\ \bibnamefont
  {Zheleznyakov}},\ }\href@noop {} {\emph {\bibinfo {title} {Electromagnetic
  waves in space plasma}}}\ (\bibinfo  {publisher} {Moscow: Nauka},\ \bibinfo
  {year} {1977})\BibitemShut {NoStop}%
\bibitem [{\citenamefont {Haug}(1975)}]{Hg75}%
  \BibitemOpen
  \bibfield  {author} {\bibinfo {author} {\bibfnamefont {E.}~\bibnamefont
  {Haug}},\ }\bibfield  {title} {\bibinfo {title} {Contribution of
  electron-electron bremsstrahlung to {\relax solar x}-radiation during
  flares},\ }\href@noop {} {\bibfield  {journal} {\bibinfo  {journal} {Sol.
  Phys.}\ }\textbf {\bibinfo {volume} {45}},\ \bibinfo {pages} {453–458}
  (\bibinfo {year} {1975})}\BibitemShut {NoStop}%
\bibitem [{\citenamefont {Kontar}\ \emph {et~al.}(2007)\citenamefont {Kontar},
  \citenamefont {Emslie} \emph {et~al.}}]{Kontar}%
  \BibitemOpen
  \bibfield  {author} {\bibinfo {author} {\bibfnamefont {E.~P.}\ \bibnamefont
  {Kontar}}, \bibinfo {author} {\bibfnamefont {A.~G.}\ \bibnamefont {Emslie}},
  \emph {et~al.},\ }\bibfield  {title} {\bibinfo {title} {Electron-electron
  bremsstrahlung emission and the inference of electron flux spectra in solar
  flares},\ }\href@noop {} {\bibfield  {journal} {\bibinfo  {journal} {ApJ}\
  }\textbf {\bibinfo {volume} {670}},\ \bibinfo {pages} {857} (\bibinfo {year}
  {2007})}\BibitemShut {NoStop}%
\bibitem [{\citenamefont {Reznikova}\ \emph {et~al.}(2009)\citenamefont
  {Reznikova}, \citenamefont {Melnikov}, \citenamefont {Shibasaki} \emph
  {et~al.}}]{Rezn}%
  \BibitemOpen
  \bibfield  {author} {\bibinfo {author} {\bibfnamefont {V.~E.}\ \bibnamefont
  {Reznikova}}, \bibinfo {author} {\bibfnamefont {V.~F.}\ \bibnamefont
  {Melnikov}}, \bibinfo {author} {\bibfnamefont {K.}~\bibnamefont {Shibasaki}},
  \emph {et~al.},\ }\bibfield  {title} {\bibinfo {title} {2002 august 24 limb
  flare loop: Dynamics of microwave brightness distribution},\ }\href@noop {}
  {\bibfield  {journal} {\bibinfo  {journal} {ApJ}\ }\textbf {\bibinfo {volume}
  {697}},\ \bibinfo {pages} {735} (\bibinfo {year} {2009})}\BibitemShut
  {NoStop}%
\bibitem [{\citenamefont {Charikov}\ \emph {et~al.}(2012)\citenamefont
  {Charikov}, \citenamefont {Melnikov},\ and\ \citenamefont
  {Kudryavtsev}}]{Char}%
  \BibitemOpen
  \bibfield  {author} {\bibinfo {author} {\bibfnamefont {Y.~E.}\ \bibnamefont
  {Charikov}}, \bibinfo {author} {\bibfnamefont {V.~F.}\ \bibnamefont
  {Melnikov}},\ and\ \bibinfo {author} {\bibfnamefont {I.}~\bibnamefont
  {Kudryavtsev}},\ }\bibfield  {title} {\bibinfo {title} {Intensity and
  polarization of the hard {\relax x}-ray radiation of solar flares at the top
  and footpoints of a magnetic loop},\ }\href@noop {} {\bibfield  {journal}
  {\bibinfo  {journal} {Geomagnetism and Aeronomy}\ }\textbf {\bibinfo {volume}
  {52}},\ \bibinfo {pages} {1021–1031} (\bibinfo {year} {2012})}\BibitemShut
  {NoStop}%
\bibitem [{\citenamefont {Zharkova}\ \emph {et~al.}(2010)\citenamefont
  {Zharkova}, \citenamefont {Kuznetsov},\ and\ \citenamefont
  {Siversky}}]{Zharkova}%
  \BibitemOpen
  \bibfield  {author} {\bibinfo {author} {\bibfnamefont {V.~V.}\ \bibnamefont
  {Zharkova}}, \bibinfo {author} {\bibfnamefont {A.~A.}\ \bibnamefont
  {Kuznetsov}},\ and\ \bibinfo {author} {\bibfnamefont {T.~V.}\ \bibnamefont
  {Siversky}},\ }\bibfield  {title} {\bibinfo {title} {Diagnostics of energetic
  electrons with anisotropic distributions in solar flares},\ }\href@noop {}
  {\bibfield  {journal} {\bibinfo  {journal} {A\&A}\ }\textbf {\bibinfo
  {volume} {512}},\ \bibinfo {pages} {A8} (\bibinfo {year} {2010})}\BibitemShut
  {NoStop}%
\bibitem [{\citenamefont {Lin}\ \emph {et~al.}(2002)\citenamefont {Lin},
  \citenamefont {Dennis}, \citenamefont {Hurford} \emph {et~al.}}]{Lin02}%
  \BibitemOpen
  \bibfield  {author} {\bibinfo {author} {\bibfnamefont {R.~P.}\ \bibnamefont
  {Lin}}, \bibinfo {author} {\bibfnamefont {B.~R.}\ \bibnamefont {Dennis}},
  \bibinfo {author} {\bibfnamefont {G.~J.}\ \bibnamefont {Hurford}}, \emph
  {et~al.},\ }\bibfield  {title} {\bibinfo {title} {The {\relax reuven ramaty
  high-energy solar spectroscopic imager (rhessi)}},\ }\href@noop {} {\bibfield
   {journal} {\bibinfo  {journal} {Sol. Phys.}\ }\textbf {\bibinfo {volume}
  {210}},\ \bibinfo {pages} {3} (\bibinfo {year} {2002})}\BibitemShut {NoStop}%
\bibitem [{\citenamefont {Hurford}\ \emph {et~al.}(2003)\citenamefont
  {Hurford}, \citenamefont {Schwartz} \emph {et~al.}}]{Hur03}%
  \BibitemOpen
  \bibfield  {author} {\bibinfo {author} {\bibfnamefont {G.~J.}\ \bibnamefont
  {Hurford}}, \bibinfo {author} {\bibfnamefont {R.~A.}\ \bibnamefont
  {Schwartz}}, \emph {et~al.},\ }\bibfield  {title} {\bibinfo {title} {First
  gamma images of a solar flare},\ }\href@noop {} {\bibfield  {journal}
  {\bibinfo  {journal} {ApJ}\ }\textbf {\bibinfo {volume} {595}},\ \bibinfo
  {pages} {L77} (\bibinfo {year} {2003})}\BibitemShut {NoStop}%
\bibitem [{\citenamefont {Haug}(975a)}]{Hg75a}%
  \BibitemOpen
  \bibfield  {author} {\bibinfo {author} {\bibfnamefont {E.}~\bibnamefont
  {Haug}},\ }\bibfield  {title} {\bibinfo {title} {Bremsstrahlung and pair
  production in the field of free electrons},\ }\href@noop {} {\bibfield
  {journal} {\bibinfo  {journal} {Zeitschrift f{\"u}r Naturforschung}\ }\textbf
  {\bibinfo {volume} {30}},\ \bibinfo {pages} {1099} (\bibinfo {year}
  {1975a})}\BibitemShut {NoStop}%
\bibitem [{\citenamefont {Massone}\ \emph {et~al.}(2004)\citenamefont
  {Massone}, \citenamefont {Emslie}, \citenamefont {Kontar} \emph
  {et~al.}}]{Massone}%
  \BibitemOpen
  \bibfield  {author} {\bibinfo {author} {\bibfnamefont {A.~M.}\ \bibnamefont
  {Massone}}, \bibinfo {author} {\bibfnamefont {A.~G.}\ \bibnamefont {Emslie}},
  \bibinfo {author} {\bibfnamefont {E.~P.}\ \bibnamefont {Kontar}}, \emph
  {et~al.},\ }\bibfield  {title} {\bibinfo {title} {Anisotropic bremsstrahlung
  emission and the form of regularized electron flux spectra in solar flares},\
  }\href@noop {} {\bibfield  {journal} {\bibinfo  {journal} {ApJ}\ }\textbf
  {\bibinfo {volume} {613}},\ \bibinfo {pages} {1233} (\bibinfo {year}
  {2004})}\BibitemShut {NoStop}%
\bibitem [{\citenamefont {Gluckstern}\ and\ \citenamefont {{\relax Hull,
  Jr.}}(1953)}]{GHull}%
  \BibitemOpen
  \bibfield  {author} {\bibinfo {author} {\bibfnamefont {R.~L.}\ \bibnamefont
  {Gluckstern}}\ and\ \bibinfo {author} {\bibfnamefont {M.~H.}\ \bibnamefont
  {{\relax Hull, Jr.}}},\ }\bibfield  {title} {\bibinfo {title} {Polarization
  dependence of the integrated bremsstrahlung cross section},\ }\href@noop {}
  {\bibfield  {journal} {\bibinfo  {journal} {Phys. Rev.}\ }\textbf {\bibinfo
  {volume} {90}},\ \bibinfo {pages} {1030} (\bibinfo {year}
  {1953})}\BibitemShut {NoStop}%
\bibitem [{\citenamefont {Hurford}\ \emph {et~al.}(2002)\citenamefont
  {Hurford}, \citenamefont {Schmahl}, \citenamefont {Schwartz} \emph
  {et~al.}}]{Hur02}%
  \BibitemOpen
  \bibfield  {author} {\bibinfo {author} {\bibfnamefont {G.~J.}\ \bibnamefont
  {Hurford}}, \bibinfo {author} {\bibfnamefont {E.~J.}\ \bibnamefont
  {Schmahl}}, \bibinfo {author} {\bibfnamefont {R.~A.}\ \bibnamefont
  {Schwartz}}, \emph {et~al.},\ }\bibfield  {title} {\bibinfo {title} {The
  rhessi imaging concept},\ }\href@noop {} {\bibfield  {journal} {\bibinfo
  {journal} {Sol. Phys.}\ }\textbf {\bibinfo {volume} {210}},\ \bibinfo {pages}
  {61} (\bibinfo {year} {2002})}\BibitemShut {NoStop}%
\bibitem [{\citenamefont {Holman}\ \emph {et~al.}(2011)\citenamefont {Holman},
  \citenamefont {Aschwanden}, \citenamefont {Aurass} \emph {et~al.}}]{Hol11}%
  \BibitemOpen
  \bibfield  {author} {\bibinfo {author} {\bibfnamefont {G.~D.}\ \bibnamefont
  {Holman}}, \bibinfo {author} {\bibfnamefont {M.~J.}\ \bibnamefont
  {Aschwanden}}, \bibinfo {author} {\bibfnamefont {H.}~\bibnamefont {Aurass}},
  \emph {et~al.},\ }\bibfield  {title} {\bibinfo {title} {Implications of
  {\relax x}-ray observations for electron acceleration and propagation in
  solar flares},\ }\href@noop {} {\bibfield  {journal} {\bibinfo  {journal}
  {Space Sci Rev.}\ }\textbf {\bibinfo {volume} {159}} (\bibinfo {year}
  {2011})}\BibitemShut {NoStop}%
\bibitem [{\citenamefont {Haug}(1997)}]{Hg97}%
  \BibitemOpen
  \bibfield  {author} {\bibinfo {author} {\bibfnamefont {E.}~\bibnamefont
  {Haug}},\ }\bibfield  {title} {\bibinfo {title} {On the use of
  nonrelativistic bremsstrahlung cross sections in astrophysics},\ }\href@noop
  {} {\bibfield  {journal} {\bibinfo  {journal} {A\&A}\ }\textbf {\bibinfo
  {volume} {326}},\ \bibinfo {pages} {417} (\bibinfo {year}
  {1997})}\BibitemShut {NoStop}%
\bibitem [{\citenamefont {Haug}(1998)}]{Hg98}%
  \BibitemOpen
  \bibfield  {author} {\bibinfo {author} {\bibfnamefont {E.}~\bibnamefont
  {Haug}},\ }\bibfield  {title} {\bibinfo {title} {Photon spectra of
  electron-electron bremsstrahlung},\ }\href@noop {} {\bibfield  {journal}
  {\bibinfo  {journal} {Sol. Phys.}\ }\textbf {\bibinfo {volume} {178}},\
  \bibinfo {pages} {341} (\bibinfo {year} {1998})}\BibitemShut {NoStop}%
\bibitem [{\citenamefont {Leach}\ and\ \citenamefont {Petrosian}(1981)}]{LP}%
  \BibitemOpen
  \bibfield  {author} {\bibinfo {author} {\bibfnamefont {J.}~\bibnamefont
  {Leach}}\ and\ \bibinfo {author} {\bibfnamefont {V.}~\bibnamefont
  {Petrosian}},\ }\bibfield  {title} {\bibinfo {title} {Impulsive phase of
  solar flares.~1: Characteristics of high energy electrons},\ }\href@noop {}
  {\bibfield  {journal} {\bibinfo  {journal} {ApJ}\ }\textbf {\bibinfo {volume}
  {251}},\ \bibinfo {pages} {781} (\bibinfo {year} {1981})}\BibitemShut
  {NoStop}%
\bibitem [{\citenamefont {Emslie}\ \emph {et~al.}(2012)\citenamefont {Emslie},
  \citenamefont {Kontar}, \citenamefont {Krucker},\ and\ \citenamefont
  {Lin}}]{Emslie}%
  \BibitemOpen
  \bibfield  {author} {\bibinfo {author} {\bibfnamefont {A.~G.}\ \bibnamefont
  {Emslie}}, \bibinfo {author} {\bibfnamefont {E.~P.}\ \bibnamefont {Kontar}},
  \bibinfo {author} {\bibfnamefont {S.}~\bibnamefont {Krucker}},\ and\ \bibinfo
  {author} {\bibfnamefont {R.~P.}\ \bibnamefont {Lin}},\ }\bibfield  {title}
  {\bibinfo {title} {Intensity and polarization of the hard {\relax x}-ray
  radiation of solar flares at the top and footpoints of a magnetic loop},\
  }\href@noop {} {\bibfield  {journal} {\bibinfo  {journal} {Geomagnetism and
  Aeronomy}\ }\textbf {\bibinfo {volume} {52}},\ \bibinfo {pages} {1021–1031}
  (\bibinfo {year} {2012})}\BibitemShut {NoStop}%
\bibitem [{\citenamefont {Holman}\ \emph {et~al.}(2003)\citenamefont {Holman},
  \citenamefont {Sui}, \citenamefont {Schwartz},\ and\ \citenamefont
  {Emslie}}]{Hol03}%
  \BibitemOpen
  \bibfield  {author} {\bibinfo {author} {\bibfnamefont {G.~D.}\ \bibnamefont
  {Holman}}, \bibinfo {author} {\bibfnamefont {L.}~\bibnamefont {Sui}},
  \bibinfo {author} {\bibfnamefont {R.~A.}\ \bibnamefont {Schwartz}},\ and\
  \bibinfo {author} {\bibfnamefont {A.~G.}\ \bibnamefont {Emslie}},\ }\bibfield
   {title} {\bibinfo {title} {Electron bremsstrahlung hard {\relax x}-ray
  spectra, electron distributions, and energetics in the 2002 july 23 solar
  flare},\ }\href@noop {} {\bibfield  {journal} {\bibinfo  {journal} {ApJ}\
  }\textbf {\bibinfo {volume} {595}},\ \bibinfo {pages} {L97} (\bibinfo {year}
  {2003})}\BibitemShut {NoStop}%
\end{thebibliography}%

\end{document}